\documentclass[12pt,a4paper]{iopart}
\usepackage{amssymb,graphicx,cite}
\eqnobysec
 
\begin{document}
   
 
\title[Interaction between bright vortex solitons]
{Mean-field model of interaction between
bright vortex solitons in 
Bose-Einstein condensates}

\author{Sadhan K. Adhikari}
\address{Instituto de F\'{\i}sica Te\'orica, Universidade Estadual
Paulista, \\ 01.405-900 S\~ao Paulo, S\~ao Paulo, Brazil}

\date{\today}

\begin{abstract}

Using the explicit numerical solution of the axially-symmetric
Gross-Pitaevskii equation we study the dynamics of interaction among
vortex solitons in a rotating matter-wave bright soliton train in a
radially trapped and axially free Bose-Einstein condensate to understand
certain features of the experiment by Strecker {\it et al.} [2002 {\it
Nature} {\bf 417} 150].  In a soliton train, solitons of opposite phase
(phase $\delta =\pi$) repel and stay apart without changing shape;  
solitons with $\delta = 0$ attract, interact and coalesce, but eventually
come out; solitons with a general $\delta $ usually repel but interact
inelastically by exchanging matter. We study and suggest future
experiments with vortex solitons.

\end{abstract}
\maketitle

\section{Introduction}
 
Solitons are  solutions of  wave equation where
localization is obtained due to a nonlinear interaction. Solitons
 have been noted in  optics \cite{0},
high-energy physics and water waves \cite{1}, and more recently in
Bose-Einstein
condensates (BEC) \cite{2,3,4}. The 
bright
solitons of BEC represent  local maxima \cite{3,4,4a0}, whereas 
dark solitons 
represent local
minima \cite{2,4a1}.  
In addition to the observation of an
isolated bright soliton in an expulsive potential
\cite{4},  a number of bright solitons constituting a
soliton train was
also  observed in an experiment  by Strecker {\it et al.}
\cite{3}, where they suddenly turned a repulsive BEC of $^7$Li
atoms attractive. 
Consequently, the BEC collapsed, exploded and 
generated a soliton train. Similar collapse and  explosion in $^{85}$Rb
BEC have been studied before from a different point of view \cite{12}.
These experiments were performed by manipulating   
the background magnetic field near a    Feshbach
resonance \cite{4a}.  It was found
\cite{3} that solitons
in such a train usually stay apart. Also, often a soliton was found
to be missing from a train \cite{3}. Hasegawa \cite{quote} considered the
generation of a train of 
optical solitons by an induced modulational instability.
There have been theoretical attempts
\cite{6,6a0,leun,sala,6a1} to simulate essentials of these experiments
\cite{3,4} on
bright
solitons in BEC.  Al Khawaja {\it et al.} \cite{6}   used a variational
approach
to describe the experiment by Strecker  {\it et al.} \cite{3} whereas
Salasnich   {\it et al.}  \cite{6a0}   and Leung  {\it et al.} \cite{leun}
used an
effective one-dimensional
model
for the same purpose. In a more recent work Salasnich   {\it et al.}
\cite{sala} considered a three-dimensional mean-field model to study the
production of and interaction between bright solitons to account for 
different aspects of the experiment by Strecker  {\it et al.} \cite{3}.
 Carr  {\it et al.}
\cite{6a1}  used an  approximate approach to describe the experiment of 
Khaykovich  {\it et al.} \cite{4}.  Elyutin  {\it et al.}  and
Shchesnovich   {\it et al.}  \cite{7x}
studied solitons in one dimension.

We use the explicit numerical solution of the
axially-symmetric  mean-field Gross-Pitaevskii (GP) equation \cite{8} to
study 
the dynamics of bright solitons in a soliton train 
\cite{3}. 
It is found that in an axially-symmetric configuration with no axial
trap, the GP equation permits 
solution in the form of a radially confined soliton train.  
We also give an explanation of missing solitons in the experiment of
Strecker  {\it et al.} \cite{3}. 

 The present approach is also  extended  
to the  study of vortex solitons  with an angular momentum of $\hbar$
per atom  in  the axial direction. Vortex solitons are  rotating 
solitons of an attractive condensate.  
Due to experimental observation
\cite{6a}  of a
vortex state  in a rotating BEC, the experimental generation of vortex
solitons seems possible. 
Attractive BEC's may not form
vortices in a thermodynamically  stable state. 
However, such vortices may be created via a Feshbach resonance \cite{4a}. 
Due to  the  conservation of angular momentum, a
vortex soliton train could be generated 
by suddenly changing the inter-atomic interaction in an axially-symmetric  
rotating vortex
condensate from repulsive to attractive near a Feshbach resonance in the
same fashion 
as in the experiment by Strecker  {\it et al.} \cite{3} for a
non-rotating BEC. Alternatively, a single vortex soliton could be prepared
and studied in the
laboratory by forming a vortex in a small repulsive condensate and then
making
the interaction attractive via a Feshbach resonance 
and subsequently  reducing the axial trap slowly. Already, there have been
theoretical considerations for these vortex states \cite{vs,9}.
It would be worthwhile 
to explore these possibilities experimentally. 

In particular we study in some detail the interaction between
two bright vortex solitons for different relative  phases between
them. Usually,
the interaction is found to be repulsive and inelastic with exchange of
particles. The interaction turns attractive for small values of relative
phase.
In one dimension the interaction is usually elastic without
exchange of particles \cite{1}.

In section 2 we present the mean-field model of solitons that we use in
the numerical analysis. In section 3 we present the results of our
numerical study. Finally, in section 4 we present the conclusions.

\section{Nonlinear Mean-free Model for Solitons}

In the following one-dimensional nonlinear free Schr\"odinger equation in
dimensionless units 
\begin{equation} \label{1}
\left[-i
 \frac{\partial }{\partial t} 
-  \frac    {\partial^2 }{\partial y^2} -
| \Psi(y,t)|^2 \right]
\Psi(y,t)  =0
\end{equation}
solitons are  bound states due to the attractive nonlinear
interaction with wave function  at time
$t$ and position
$y$:  $\Psi(y,t)= \sqrt{2|\Omega| }\exp(-i\Omega t){\mbox{sech}} (y\sqrt
{|\Omega|})$, with $\Omega$ the energy \cite{7}. 
Equation  (1) can sustain any
number of such solitons at different positions. Although,
the 
one-soliton solution of  (1) may remain stationary at a fixed position, the
many solitons of a multiple-soliton solution of this equation generally
move around because of the interaction among them \cite{5}.

The time-dependent Bose-Einstein condensate wave
function $\Psi({\bf r};\tau)$ at position ${\bf r}$ and time $\tau $
is described by the following  mean-field nonlinear GP equation
\cite{8}
\begin{eqnarray}\label{a} \left[- i\hbar\frac{\partial
}{\partial \tau}
-\frac{\hbar^2\nabla^2   }{2m}
+ V({\bf r})
+ gN|\Psi({\bf
r};\tau)|^2
 \right]\Psi({\bf r};\tau)=0,
\end{eqnarray}
where $m$
is
the mass and  $N$ the number of atoms in the
condensate,
 $g=4\pi \hbar^2 a/m $ the strength of inter-atomic interaction, with
$a$ the atomic scattering length.  
 For an axial trap    $  V({\bf
r}) =\frac{1}{2}m \omega ^2(r^2+\lambda^2 z^2)$ where
 $\omega$ is the angular frequency
in the radial direction $r$ and
$\lambda \omega$ that in  the
axial direction $z$, with $\lambda$ the aspect ratio. 
The normalization condition  is
$ \int d{\bf r} |\Psi({\bf r};\tau)|^2 = 1. $

In  a  quantized  vortex state \cite{9}, with   each atom having  
angular momentum $L\hbar$ along the $z$ axis,
$\Psi({\bf r}, \tau)= \psi(r,z,\tau)\exp (iL\theta)
$ where  $\theta$ is the azimuthal
angle.
Now  transforming to
dimensionless variables $x =\sqrt 2 r/l$,  $y=\sqrt 2 z/l$,   $t=\tau \omega, $
$l\equiv \sqrt {\hbar/(m\omega)}$,
and
${ \varphi(x,y;t)} \equiv   x\sqrt{{l^3}/{\sqrt
8}}\psi(r,z;\tau),$   (\ref{a}) becomes \cite{9}
\begin{eqnarray}\label{d1}
&\biggr[&-i\frac{\partial
}{\partial t} -\frac{\partial^2}{\partial
x^2}+\frac{1}{x}\frac{\partial}{\partial x} -\frac{\partial^2}{\partial
y^2}
+\frac{1}{4}\left(x^2+\lambda^2 y^2\right) \nonumber \\
&+& {L^2-1\over x^2}  +                                                          
8\sqrt 2 \pi n\left|\frac {\varphi({x,y};t)}{x}\right|^2
 \biggr]\varphi({ x,y};t)=0, 
\end{eqnarray}
where non-linearity 
$ n =   N a /l$. For solitonic states $n$ is negative.
In terms of the 
one-dimensional probability 
$P(y,t)$ defined by 
\begin{equation}\label{avi}
P(y,t) = 2\pi \int_0 ^\infty 
dx |\varphi(x,y,t)|^2/x , 
\end{equation}
the normalization of the wave function 
is given by $\int_{-\infty}^\infty dy P(y,t) = 1$

We solve the GP equation (\ref{d1}) numerically  using a variation of the 
split-step time-iteration
method
using the Crank-Nicholson discretization scheme described recently
\cite{11}. Typical space and time steps for discretization are 0.1 and
0.001. 
 The variation of the standard approach is required for $\lambda =0$. 
The time iteration is started with the known harmonic oscillator solution
for a small $\lambda 
\equiv \lambda _0 \approx 0.05$ and nonlinearity $n=0$: $\varphi(x,y) =
[\lambda
/\{2
^{2L+3}\pi^3 (L!)^2 \}]^{1/4}$
$x^{1+L}e^{-(x^2+\lambda y ^2)/4}$ with energy $(1+L+\lambda/2)$
\cite{9}. The desired 
value of nonlinearity $n$ is then slowly switched on and  $ \lambda  $
slowly switched off from  $ \lambda = \lambda _0 $ to zero
in the course of
time iteration.
The
solution then corresponds to the trapped BEC for $ \lambda  =0$. 
Then, without changing any
parameter, this solution is iterated several thousand  times so that a
solution  for  $\lambda = 0$ is obtained 
independent of the initial input
$\lambda =\lambda _0$.

\section{Numerical Results}

A classic soliton  in three-dimensional BEC 
 can be realized for attractive  nonlinear  potential
($n<0$) by setting $\lambda =0$  \cite{4a0} in  (\ref{d1}) . For  a
fixed
$L$, the
BEC is then governed by the  single parameter $n$. The absence of a
trap in the axial $y$ direction will allow a free movement of  the
solitons in this direction. Consequently, the  study of soliton
interaction will be trap independent.  
A localized BEC so created should
be the three-dimensional analogue of the one-dimensional soliton. 

However,
for calculational or experimental convenience, in the recent studies some
weak potential was applied in the axial $y$ direction. 
  In the classic
experiment
of Strecker {\it et al.} \cite{3} an optical trap was maintained in the
$y$
direction. In the recent theoretical investigation   by
Al Khawaja {\it et al.} \cite{6} 
on this experiment 
a weak harmonic trap 
was applied in this direction, 
whereas Salasnich {\it et al.} \cite{6a0} employed infinite walls 
in an  effective
one-dimensional model.
 In the
experiment
by Khaykovich {\it et al.} \cite{4} and in the related theoretical study
by Carr  {\it et al.} \cite{6a1}
an expulsive potential
($\lambda^2<0$) was
applied in  $y$
direction. 
We note that for 
$\lambda^2<0$, only {\it meta-stable} and no stable soliton  of
 (\ref{d1}) 
is possible as the potentials in this equation including the
attractive nonlinear term do not provide confinement \cite{6a1}. 
For the same reason  no bright soliton can be generated for $n\ge
0$ in
 (\ref{d1}) (repulsive condensate).

Although, under the conditions $n<0$  and   $\lambda = 0$  the potentials
of  (\ref{d1}) lead to  confinement, a soliton-type BEC state
can be generated only for $n$ greater than a critical value 
($n_{\mbox{cr}}$):  $n_{\mbox{cr}}< n<0$. For $n<n_{\mbox{cr}}$,
the system becomes too attractive and collapses and no stable soliton
could be generated. 
The collapse was first confirmed in the pioneering experiment by Gerton
{\it et al.}  \cite{10} for a
trapped BEC of $^7$Li.  The actual value of $n_{\mbox{cr}}$ is a function
of the trap parameter $\lambda$. For the spherically symmetric case
$\lambda = 1$, and  $n_{\mbox{cr}}= -0.575$ \cite{8,9}.  The value of
$n_{\mbox{cr}}$ should
slightly change for  $\lambda=0$.

\begin{figure}
 
\begin{center}

\includegraphics[width=0.49\linewidth]{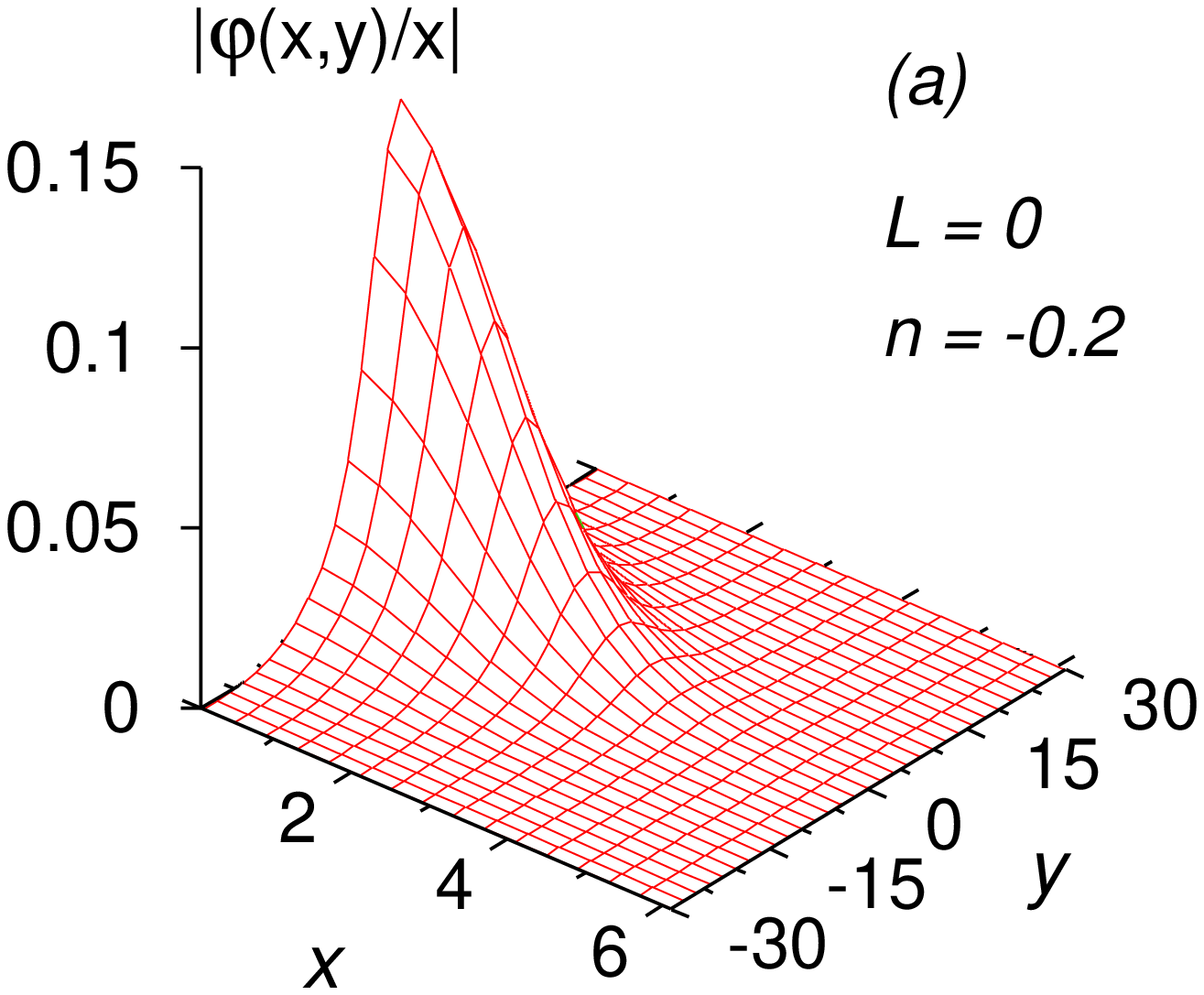}
\includegraphics[width=0.49\linewidth]{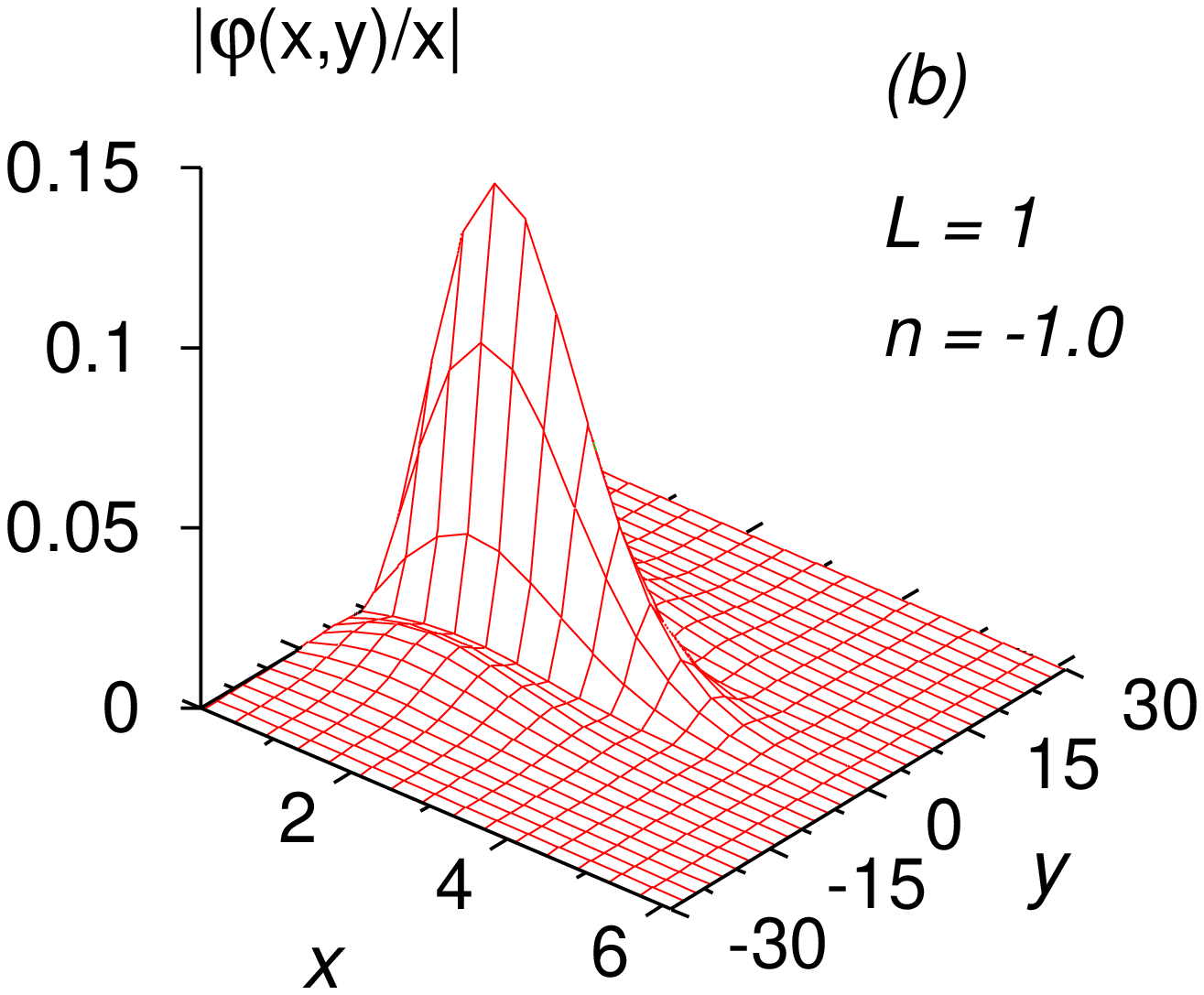}
\end{center}
 
\caption{Three-dimensional  wave function  $|\varphi
(x,y)/x|$
vs. $x$ and $y$
for a single soliton  with $\lambda
= 0$, and   (a) $L=0$, $n=-0.2$, and (b) $L=1$, $n=-1$. } 
\end{figure}

We solve the GP equation  for solitons with $\lambda =0$
and $L=0 $ and 1 \cite{9}. We find numerically that the
critical
$n$ for collapse of a single soliton is
$n_{\mbox{cr}}= -0.67$ for $L=0$ and  
 $n_{\mbox{cr}}=  -2.10$ for $L=1$. 
The result for $L=0$ is in good 
agreement with (a) the numerical result $-0.676$  obtained by 
Gammal {\it et al.} \cite{crit} using the Crank-Nicholson method, 
(b) the numerical result  $-0.676$  obtained by  P\'erez-Garc\'ia  {\it et
al.} \cite{4a0} using the steepest-descent method to minimize the
mean-field
Hamiltonian [these authors quote $Q=-8\pi Na/l =17$ instead of $n=Na/l$] 
and
(c) the result $-2/3$ obtained by Salasnich {\it et al.}
\cite{6a0} using an approximate analytic one-dimensional model.
 By solving the
three-dimensional GP equation Salasnich {\it et al.} \cite{6a0} found
their
result to be very accurate. 
However, there is some discrepancy between these results for $L=0$ 
and  the value   $n_{\mbox{cr}}=-0.6268\pm
0.0035$ obtained by 
Carr and Castin
\cite{6a1} 
using 
the
imaginary time relaxation method.  
Further independent studies
are necessary  to resolve the discrepancy. For
$\omega = 2\pi\times 800$ Hz
and final scattering length $-3a_0$
as in the experiment of Strecker {\it et al.} \cite{3},  $n_{\mbox{cr}}=
-0.67$ corresponds to about 6000 $^7$Li atoms. One can have
proportionately about
three times more atoms in the $L=1$ state.

A $L=0$    soliton with $n=-0.2$ is illustrated in figure 1
(a) where we plot the three-dimensional  wave function  $|\varphi
(x,y)/x|$
 vs. $x$  and $y$.   Next we consider $L=1$. 
In this case we calculated the soliton for $n=-1$
and plot the three-dimensional  wave function  $|\varphi
(x,y)/x|$
  in figure 1 (b). 
Because of the radial trap the soliton remains confined in the radial
direction  $x$, although free to move in the axial $y$ direction. 
In either case  the single
soliton remains  stable for more than
400 000 time iterations.

\begin{figure}
 
\begin{center}
\includegraphics[width=0.49\linewidth]{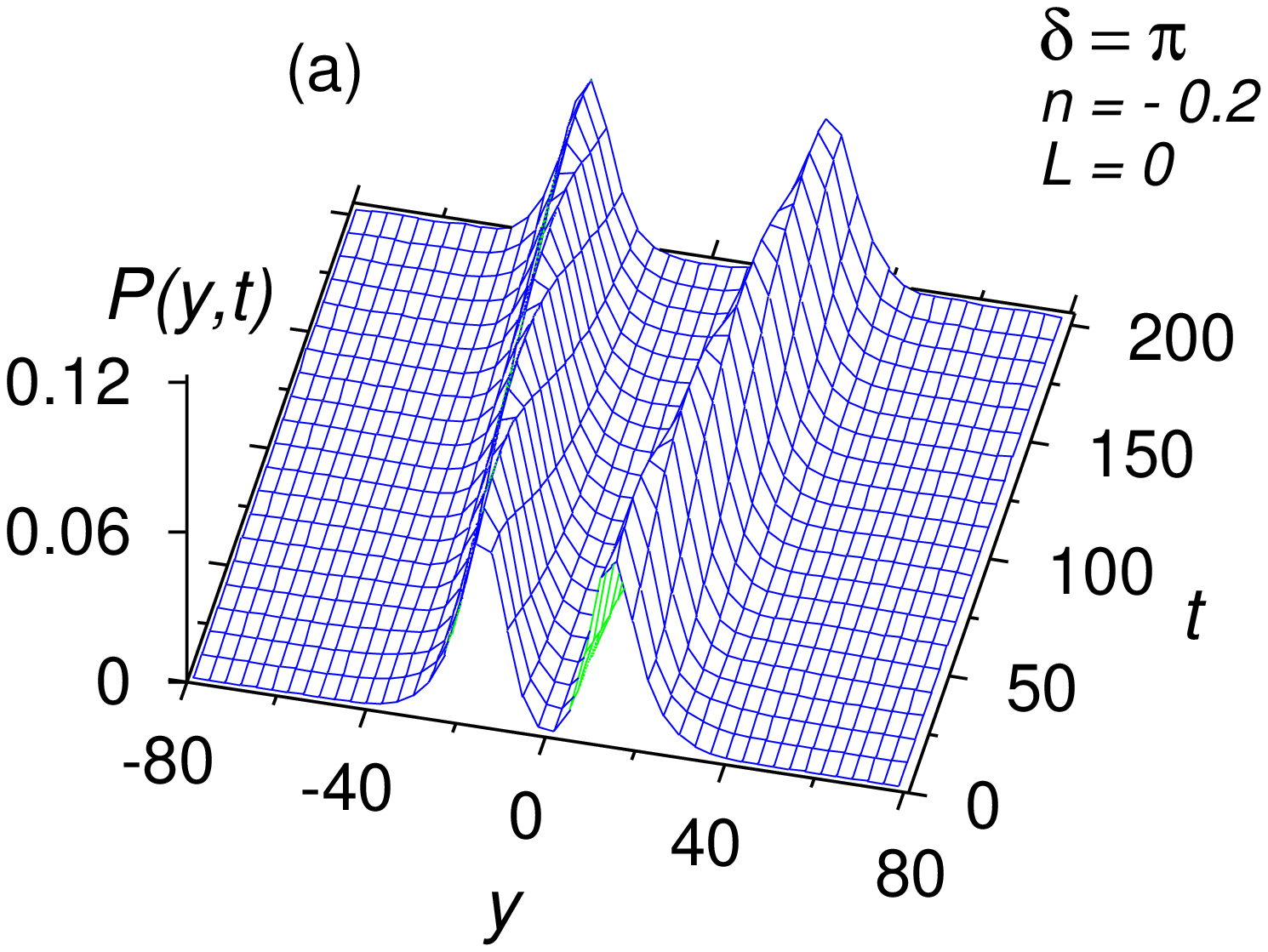}
\includegraphics[width=0.49\linewidth]{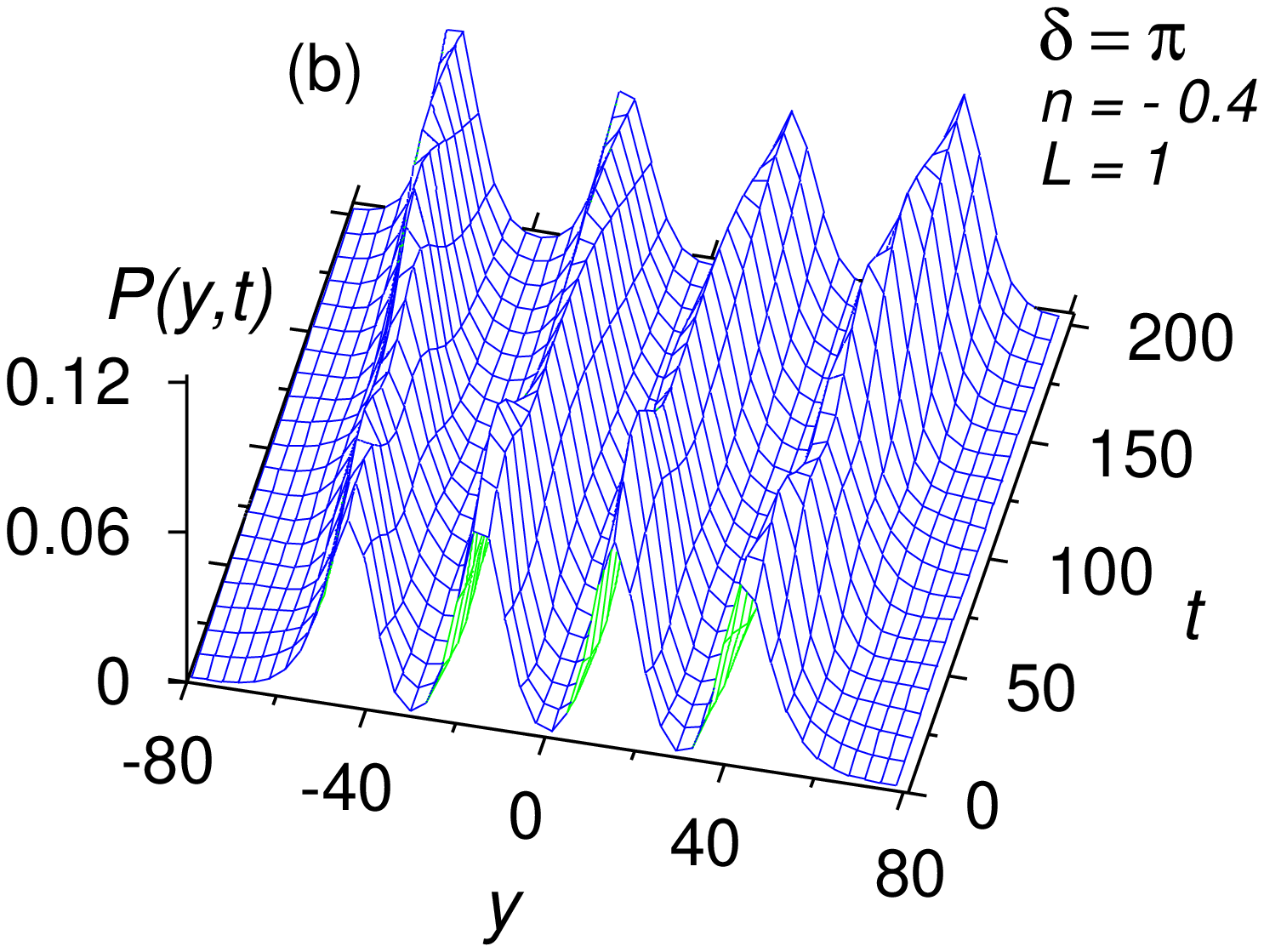}
\end{center}
 
\caption{One-dimensional probability $P(y,t)$
vs. $y$ and $t$ for a train  of (a) two solitons each with 
$n=-0.2$, $\delta=\pi$,  $L=0$,   and (b) four $L=1$ vortex 
solitons each 
with $n=-0.4$ and a phase difference 
 $\delta=\pi$
between neighboring solitons.}
\end{figure}

After having demonstrated the formation of a single soliton we next
consider the dynamics of two solitons in a soliton train. 
Two solitons are 
prepared at positions $y_1$ and $y_2$  and then superposed with a phase
difference $\delta$. Specifically, we consider the following  
superposition of two normalized solitonic waves $\bar \varphi $ at $y=\pm
y_0$
with phase
difference $\delta$ between them 
\begin{equation}\label{xxx}
\varphi(x,y)=|\bar \varphi(x,y+y_0)|+ e^{i\delta}  
|\bar \varphi(x,y-y_0)|,
\end{equation}
with $2y_0$ the initial separation between the solitons 
in the axial direction. 
To conserve the total number of atoms we normalize the
superposed wave function
(\ref{xxx}) of the two solitons  to 2 as it contains twice as many
particles
as in a single soliton.  
In this fashion one can also construct the
superposition of several solitons with a specific phase difference between
them.  
The time evolution of the  soliton train so formed upon
superposition is found using the iterative solution of  (\ref{d1}).
We present results for a soliton train  with  two vortex solitons of 
same nonlinearity with a phase
difference $\delta$ of $\pi, 3\pi/4,$ $\pi/2$, $\pi/4$,  $\pi/8$ and 0
between
them. Consideration of two equal solitons does not lead to a
specialization and  we shall
see that for a general $\delta$ two equal initial solitons generally  lead
to two
unequal solitons due to exchange of atoms.

In one dimension the solitons attract for $\delta =0$ and repel for
$\delta = \pi$ \cite{5,6}. In the present three-dimensional case
for a general $\delta$ a more complicated
motion emerges.  In the repulsive  $\delta =\pi $ case,  two
solitons stay away from each other. In the attractive  $\delta =
0$ case, they come close, coalesce, interact and come out.
 In a
one-dimensional model of two trapped ($\lambda \ne 0$) three-dimensional
solitons, it has been shown that these solitons repel for $\delta = \pi$
\cite{6,6a0}. In the following we study the rich dynamics of soliton
interaction in a bright soliton train in three dimensions. The solitons
can easily be set 
into motion by applying a perturbation or a constant (gravitational) force
in $y$ direction. However, for calculational convenience we chose not to
do that and thus  we studied the interesting relative motion between
solitons suppressing 
their center-of-mass  motion.  In all the cases studied the initial 
velocity of the solitons is zero. 
The numerical simulation was performed on a lattice 
$7 >  x\ge 0$ and $250>y>-250$.

We studied the dynamics of interaction of two solitons  in
view of the experiment by Strecker {\it et al.} \cite{3}.  For simplicity
first
we prepared two solitons of equal mass corresponding to the same
nonlinearity $n=-0.2$ centered at points $ y = \pm 15$ with angular
momentum
$L =0 $ and introduced them as the input to the GP equation with the
initial phase difference $\delta= \pi$. The time evolution of the train of
two such solitons is
exhibited in figure 2 (a) where we plot the one-dimensional probability
$P(y,t)$ of (\ref{avi}) vs.
$y$ and $t$. Because of mutual repulsion for $\delta=\pi$,
the two solitons stay apart and move away from each other. 
In an
interval of time 200, the two solitons moved from positions $y=\pm 
15$ to $\pm 25$, respectively. 

\begin{figure}[!ht]
\begin{center}
\includegraphics[width=0.49\linewidth]{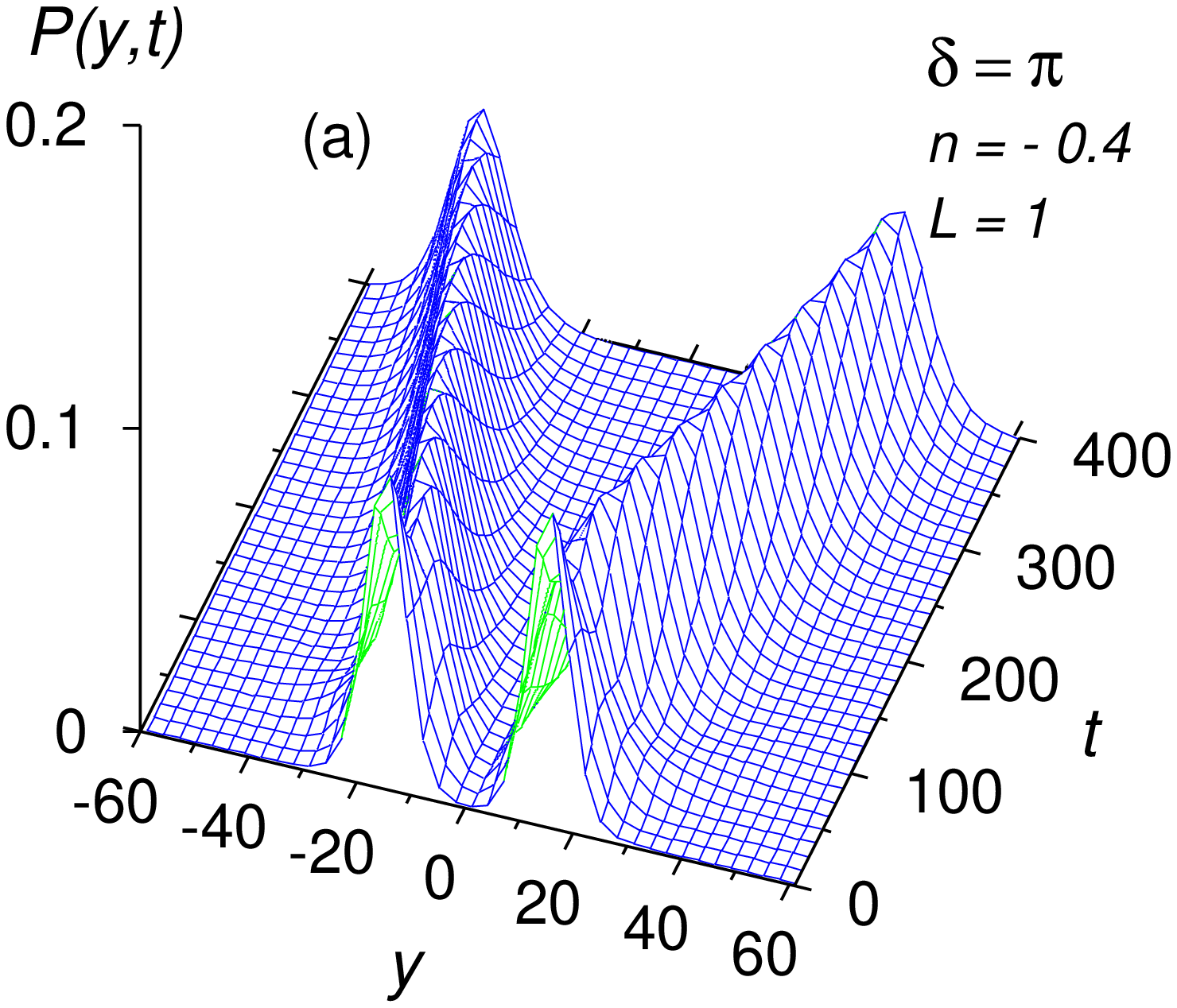}
\includegraphics[width=0.49\linewidth]{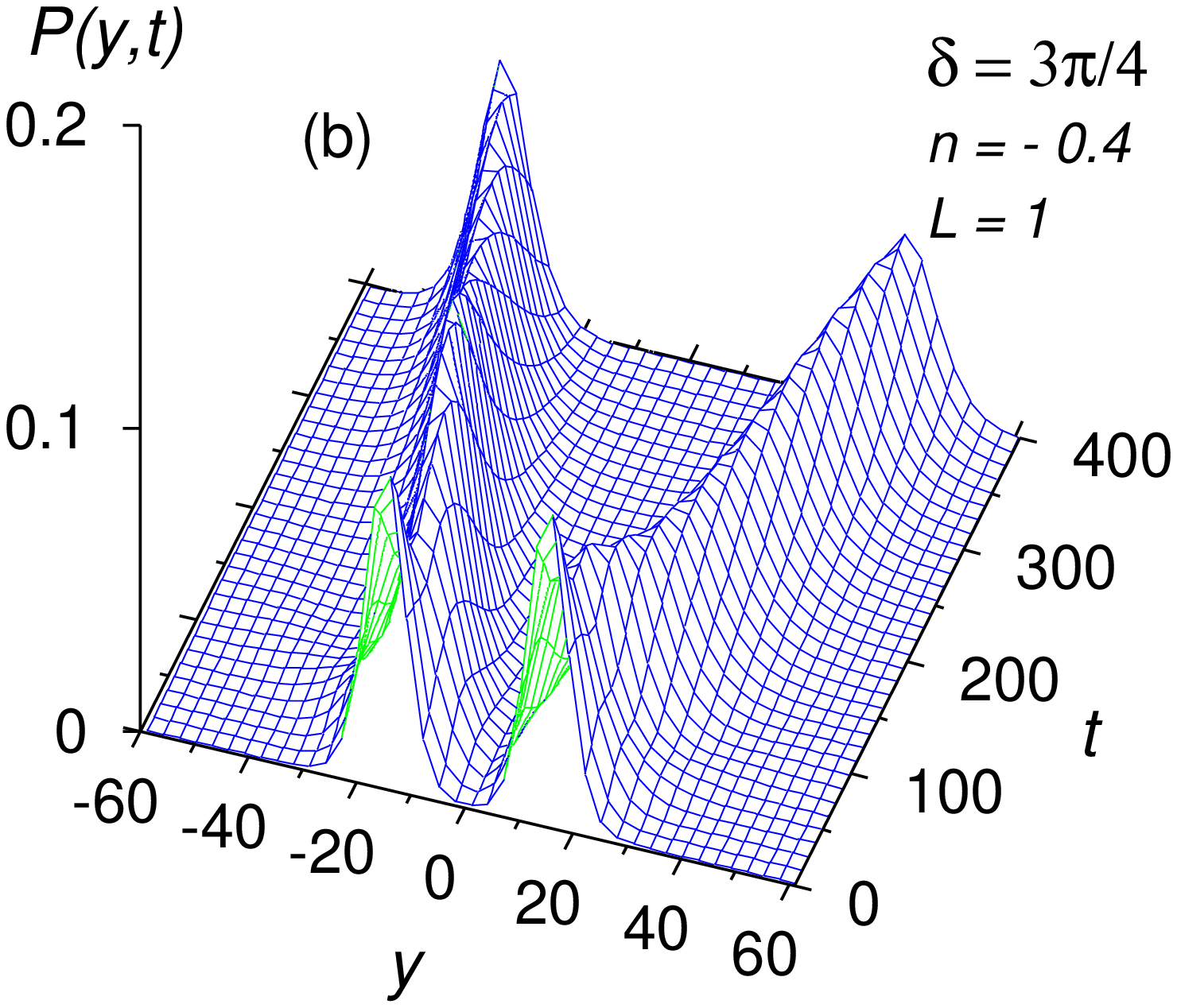}
\includegraphics[width=0.49\linewidth]{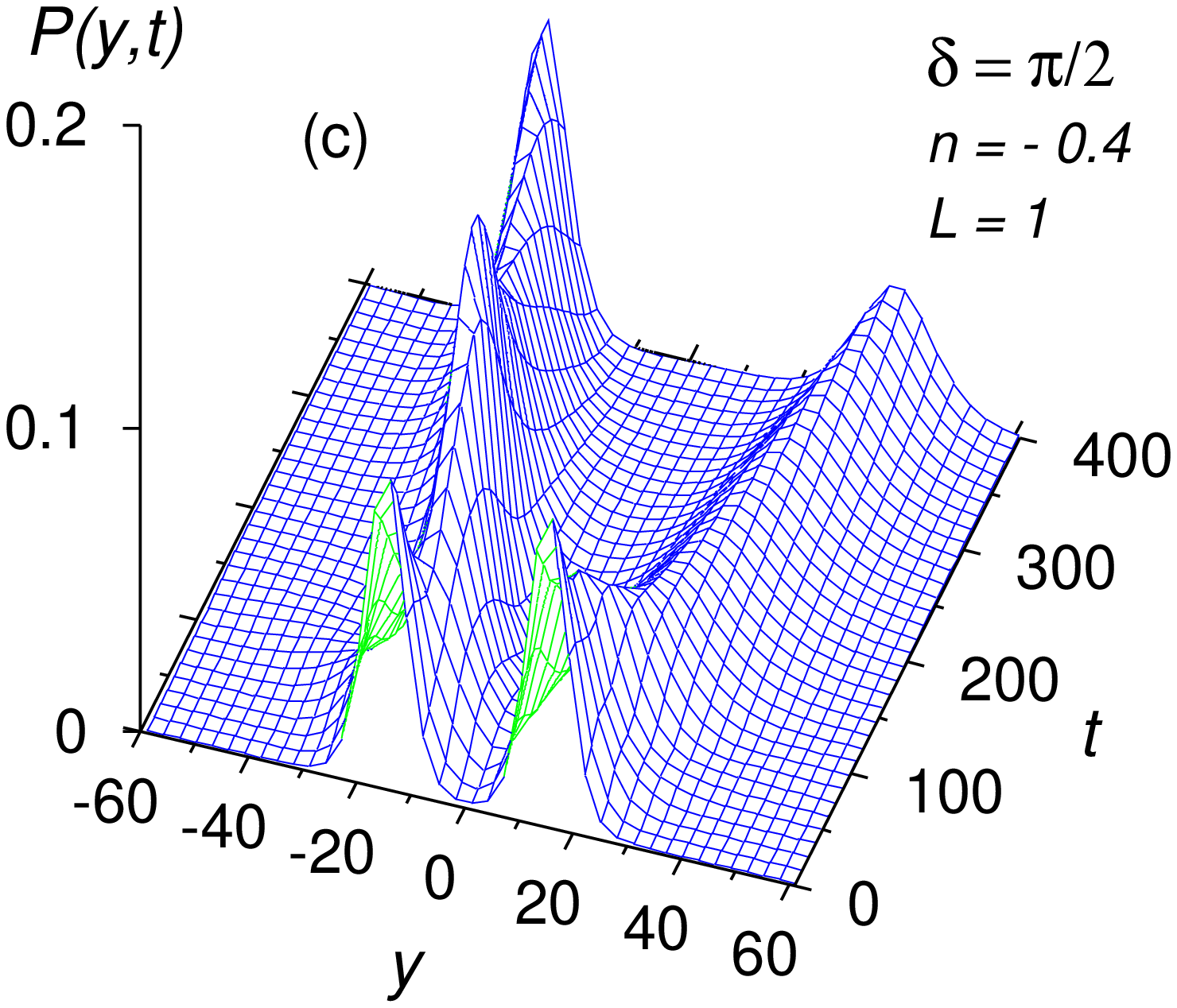}
\includegraphics[width=0.49\linewidth]{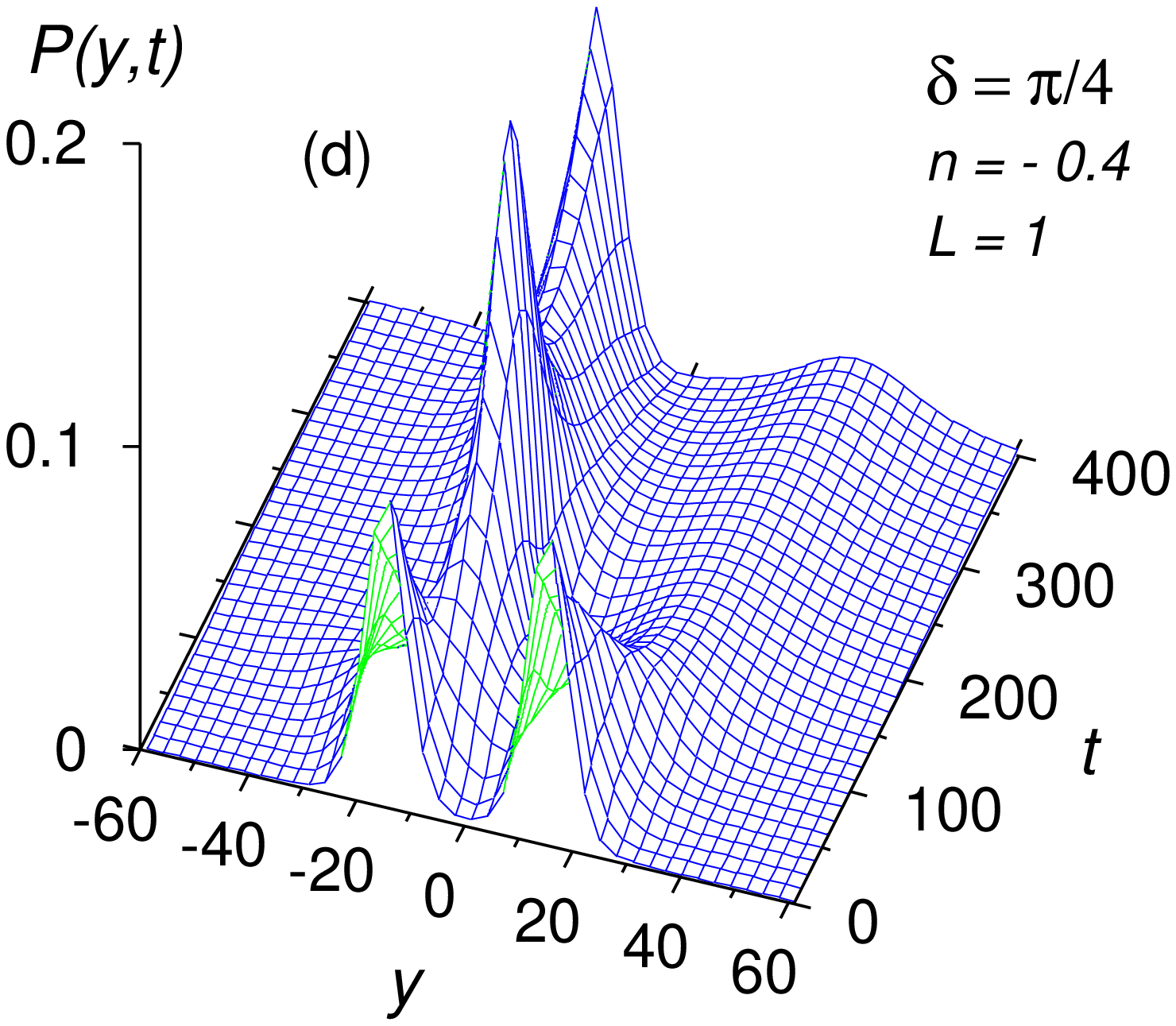}
\includegraphics[width=0.49\linewidth]{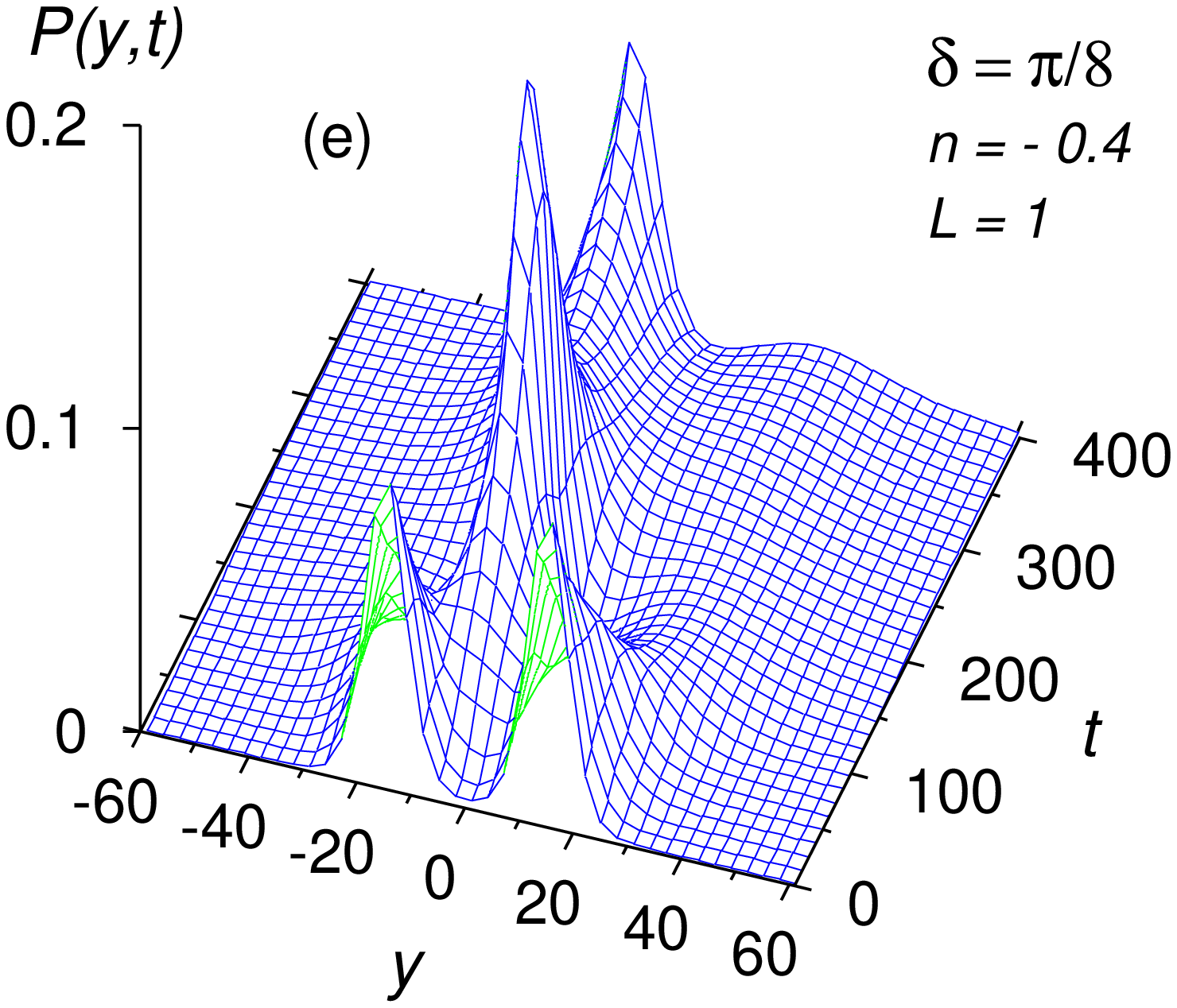}
\includegraphics[width=0.49\linewidth]{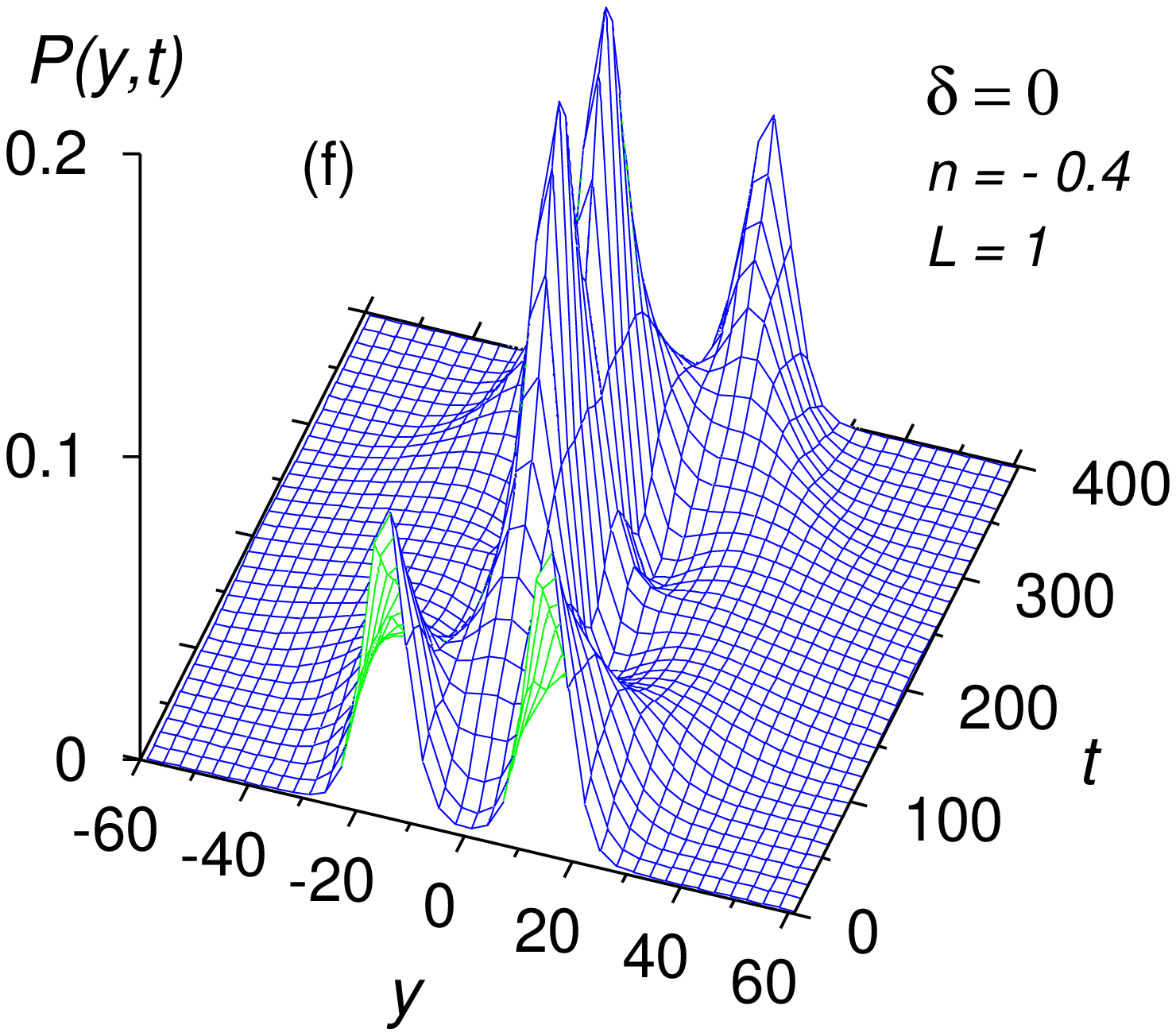}
\end{center}

\caption{One-dimensional probability  $P(y,t)$
vs. $y$ and $t$ for a train of two vortex solitons each with  $n = -0.4,
L=1$ and $\delta =$\-  (a) $\pi $,
(b) $3
\pi /4$,
(c)  $ \pi/2$ and (d) $\pi/4$, (e) $\pi/8$ and (f) 0.}

\end{figure}

We studied soliton trains with both $L=0$ and $L=1$ solitons and they lead
to physically similar results. In the rest of the study we consider only
$L=1$ vortex solitons as they were never studied before.  Also, an
effective one-dimensional model usually employed in other studies does not
seem to be applicable to the study of a vortex soliton. The rotational
degree of freedom responsible for the generation of vortex does not exist
in a one-dimensional model.  We consider
four
$L=1$ vortex solitons each  with $n= -0.4$ at positions $\pm 45$ and $\pm
15$
with
phase difference $\delta =\pi$ between two neighboring and hence,
repelling, solitons. Obviously, one can accommodate as many 
repelling solitons 
as one prefers in a train. Because of the
repulsion, the solitons  stay apart and move forward without interaction
and
maintaining shape. Such a soliton train is illustrated in figure 2 (b). In
an interval of time 200, the four solitons have moved from positions
$y=\pm 15$ and $\pm 45$ to $\pm 17$ and $ \pm 52$, respectively.

Now we consider the interaction between two vortex solitons in some
detail. 
For that  we reduced the  phase  $\delta$
between two $L=1$ vortex solitons at $y=\pm 15$ each with $n= -0.4$
from $\pi$ to 0. In figures 3 (a), (b), (c), (d), (e) and (f) we plot
one-dimensional
probability $P(y,t)$  of (\ref{avi})  for two interacting vortex solitons
vs. $y$ and $t$ for $\delta = \pi,$ $3\pi/4, \pi
/2, \pi/4, \pi/8$ and 0,
respectively. 
As $\delta$ is reduced from
$\pi$ to $\pi/2$ the two solitons continue to repel and 
stay apart without
much change in shape. However, while
 $\delta$ is reduced from 
$\pi$  the
repulsive 
interaction between the solitons gradually become ``inelastic" and there
is exchange
of atoms  between the solitons.  
Because of exchange of atoms 
the motion of
solitons in figures 3 is asymmetric in general except for $\delta =\pi$
and 0.

In figure 3 (a) for $\delta=\pi$ in the interval of time 400, the solitons
moved from $y = \pm 15$ to $\pm 38$, respectively, corresponding to a
final separation between the two solitons of 76. In figure 3 (b)  for
$\delta=3\pi/4$ in the interval of time 400, the solitons moved from $y =
\pm 15$ to 35 and $-41$, respectively, again leading to a final separation
of 76. Although the repulsion is same in both cases, an asymmetry has
appeared in figure 3 (b) due to inelastic collision with exchange of
atoms.   In figure 3 (c) for $\delta = \pi/2$ in the
interval of time 400, the solitons moved from $y = \pm 15$ to 38 and
$-26$, respectively, corresponding to a final separation of 64. As
$\delta$ is reduced further the similar trend is maintained:  reduction in
repulsion due to inelasticity (exchange of atoms between the two
solitons).  In figure 3 (d) we see that for $\delta =\pi/4$ the two
solitons maintain their identity, exchange atoms and the overall
interaction continues to be repulsive. During the interval of time 400, in
figure 3 (d) the solitons moved from $y=\pm 15$ to 33 and $-16$,
respectively, corresponding to a final separation of 49.  For a further
reduction of $\delta$ from $\pi/4$ the inelasticity and asymmetry of
interaction reduces.  The mutual overall interaction becomes less and less
repulsive which finally becomes attractive as $\delta \to 0$. In figure 3
(e) we show the relative motion of two solitons for $\delta =\pi/8$. The
two solitons have a tendency to combine to form a single soliton before
they lead to two separate solitons. In time 400, solitons moved from $y
=\pm 15$ to 22 and $-11$, respectively, corresponding to a separation of
33. The relative movement between the solitons has reduced monotonically
as $\delta$ is changed from $\pi$ to $\pi/8$.  From figure 3 (f) we see
that for $\delta = 0$ there is strong overall attraction and the two
solitons interact and coalesce to form a single soliton which eventually
gives birth to two solitons of same shape as the initial ones. The final
separation between the two solitons in figures 3 (a), (b), (c), (d) and
(e) for $\delta= \pi, 3\pi/4, \pi/2, \pi/4$ and $\pi/8$, respectively, are
76, 76, 64, 49 and 33 showing a gradual reduction in repulsion as $\delta$
is reduced from $\pi$ to $\pi/8$. In all cases the final separation is
larger than the initial separation $2y_0 (=30)$ between solitons
reflecting an overall repulsion.  The asymmetry in the position of the
two final solitions is zero for $\delta=0$ and $\pi$ and largest for
$\delta=\pi/4$.

\begin{figure}
 
\begin{center}
\includegraphics[width=0.6\linewidth]{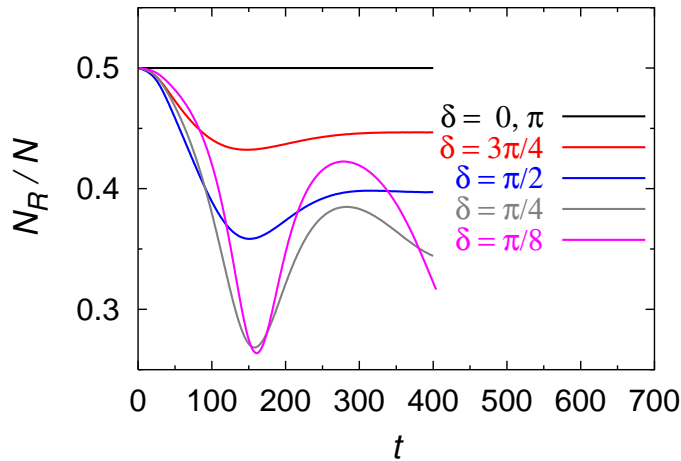}
\end{center}

\caption{Ratio  $N_R/N$ vs. $t$ for different  initial phase  
$\delta$ in Eq. (\ref{xxx}) for a soliton-train of two $L=1$
vortex solitions initially at $y=y_0=\pm 15$. 
 $N_R$ is the number of atoms in the right soliton 
at $y=+15$ at $t=0$ and $N$ the total number of atoms. 
}
 
\end{figure}

Two features of figures 3 deserve some comments. First, the value of the
phase $\delta $ for which the interaction between two solitons is
attractive. In the one-dimensional model study by Salasnich {\it et al.}
\cite{6a0} it is attractive for all $\cos \delta >0$ in agreement with a
previous analysis \cite{gord} whereas in the present three-dimensional
study with an additional radial trap there is no sharp transition from
attraction to repulsion for a fixed $\delta$.  The presence of the radial
trap in our study makes a slow transition from repulsion to attraction
with the change of $\delta$. However, there is clear attraction for
$\delta \approx 0$ and repulsion for $\cos \delta < 0$ in both models.  
Further studies are needed for a complete understanding of the present
three-dimensional solitons under transverse confinement. This transverse
confinement is absent in the one-dimensional model of Salasnich {\it et
al.} \cite{6a0}. It is true that in the one-dimensional model of Salasnich
{\it et al.} the radial trap frequency enters as a parameter [see their
equation (1)], it does not include real three-dimensional nonlinear
dynamics. The present three-dimensional solitons cannot be considered to
be classic one-dimensional integrable text-book solitons of the type
considered in references \cite{6a0,gord}. Hence the text-book wisdom is
not directly applicable to the present analysis.

The second feature is the asymmetry in figures 3 for an arbitrary $\delta$
because of inelastic exchange of atoms between the two solitons which is
absent in the one-dimensional model \cite{6a0}.  Although we report the
results for angular momentum $L=1$ in this study we have verified that
similar asymmetry is present for $L=0$.  The asymmetry is also
present in the analysis of \cite{sala} of two interacting solitons. One of
the two equal solitons increases and the other decreases in size due to
inelastic exchange of atoms. In figures 3 the left soliton grows in size
and moves slower, the right soliton becomes smaller in size and moves
faster.  The direction of particle exchange is determined by the way the
phase is introduced between the two solitons in (\ref{xxx}). We verified
that as $\delta$ is changed to $-\delta$ the role of asymmetry is
reversed. For a positive $\delta$ due to particle exchange in figures 3
the left soliton grows in size. For a negative $\delta$ the right soliton
grows in size in a symmetric fashion so that $P(y,t, -\delta)= P(-y,t,
\delta).$ The asymmetry of interaction is absent in figures 2 and 3 (a)
where the interaction is purely repulsive with no exchange of atoms.

To study systematically 
the
asymmetry of the interaction beween two solitons 
we plot in figure 4 the ratio  $N_R/N$ as a
function of time for the cases presented in figures 3. Here $N_R$ is the
number of atoms in the right  soliton and $N$ the total number in two
solitons. The variation of  $N_R/N$ is quite similar to that found  in 
\cite{sala}. For $\delta = \pi/4$ in both studies  $N_R/N$ first attains a
minimum and then oscillates, however remaining less than 0.5. 
A similar behavior is noted in the present study  for $\delta =\pi/8$.
For  $\delta> \pi/4, $   $N_R/N$ attains a minimum and then
increases a little to attain a constant value smaller than 0.5. 
However, a quantitative comparison between the two studies is not to the
point as the initial conditions of the two studies are different, e. g.,
the 
initial size and angular momentum $L$  and the initial separation between
the solitons. The details of interaction dynamics should depend on these
initial conditions. In the  present simulation we find that, 
for the change $\delta \to -\delta$,
 $N_R/N  \to  N_L/N$,  where $N_L\equiv (N-N_R)$ is the number of atoms in
the left soliton.

\begin{figure}[!ht]
\begin{center}
\includegraphics[width=0.4\linewidth]{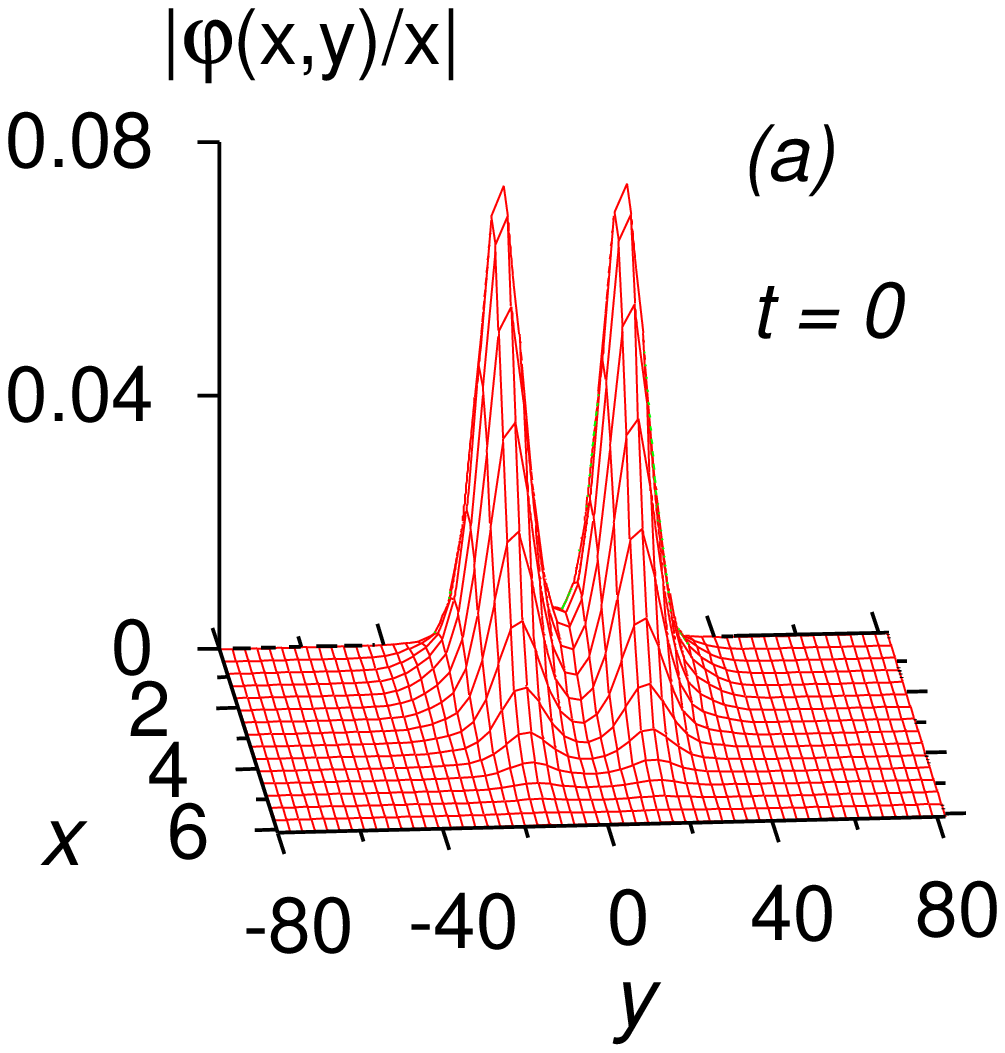}
\includegraphics[width=0.4\linewidth]{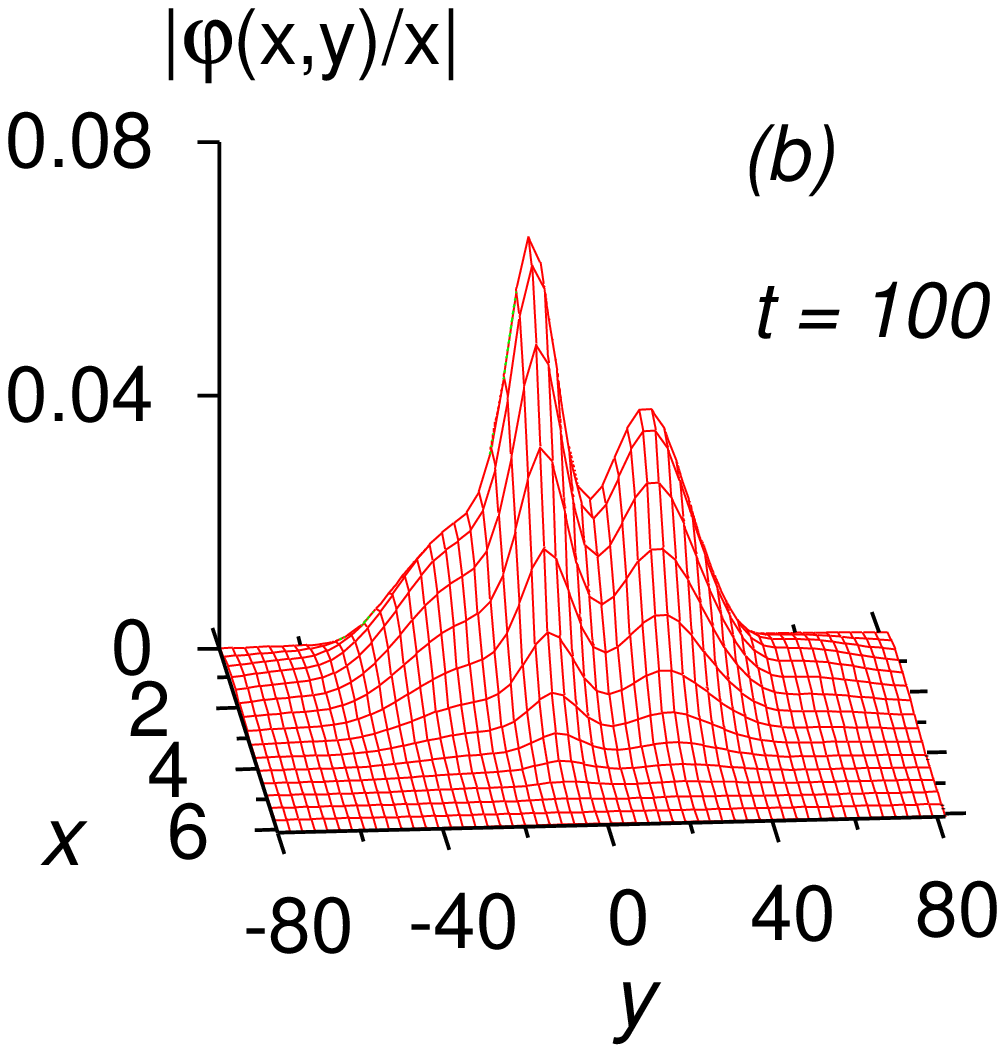}
\includegraphics[width=0.4\linewidth]{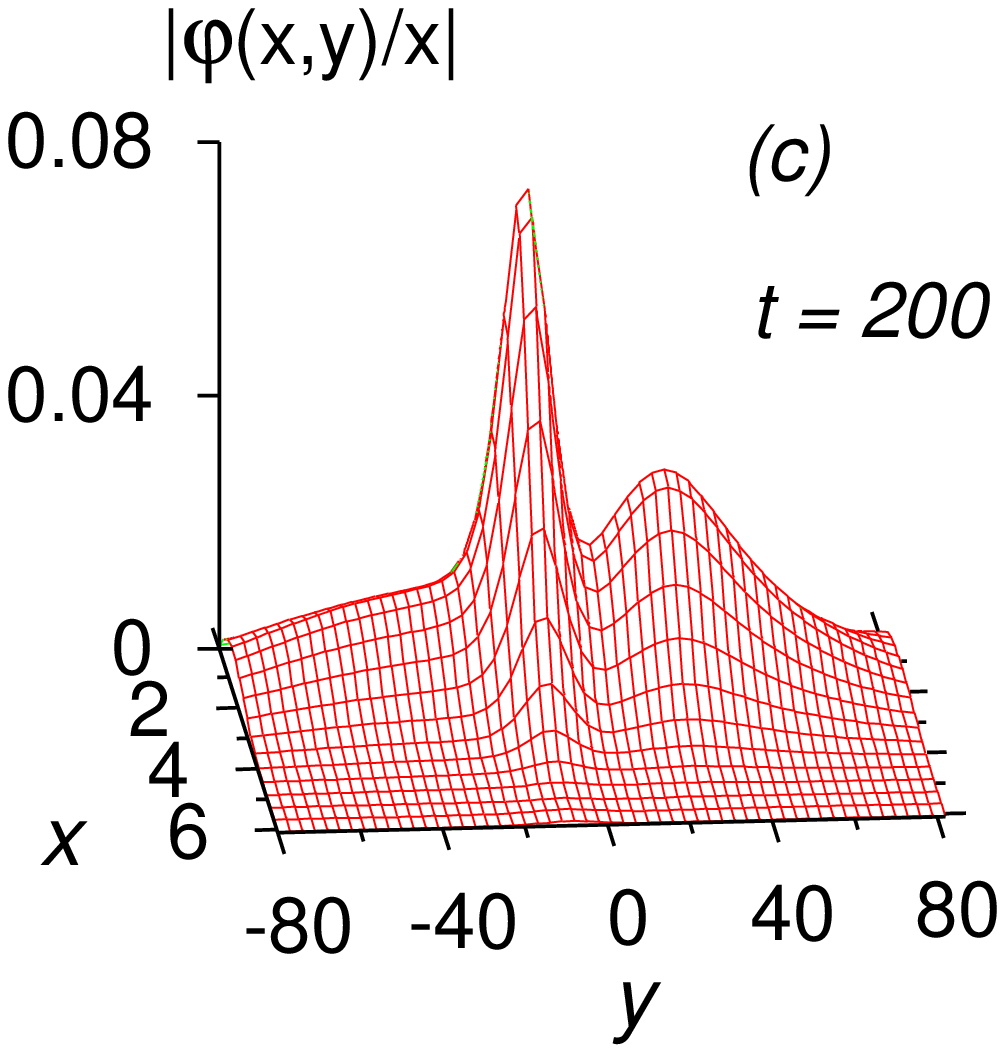}
\includegraphics[width=0.4\linewidth]{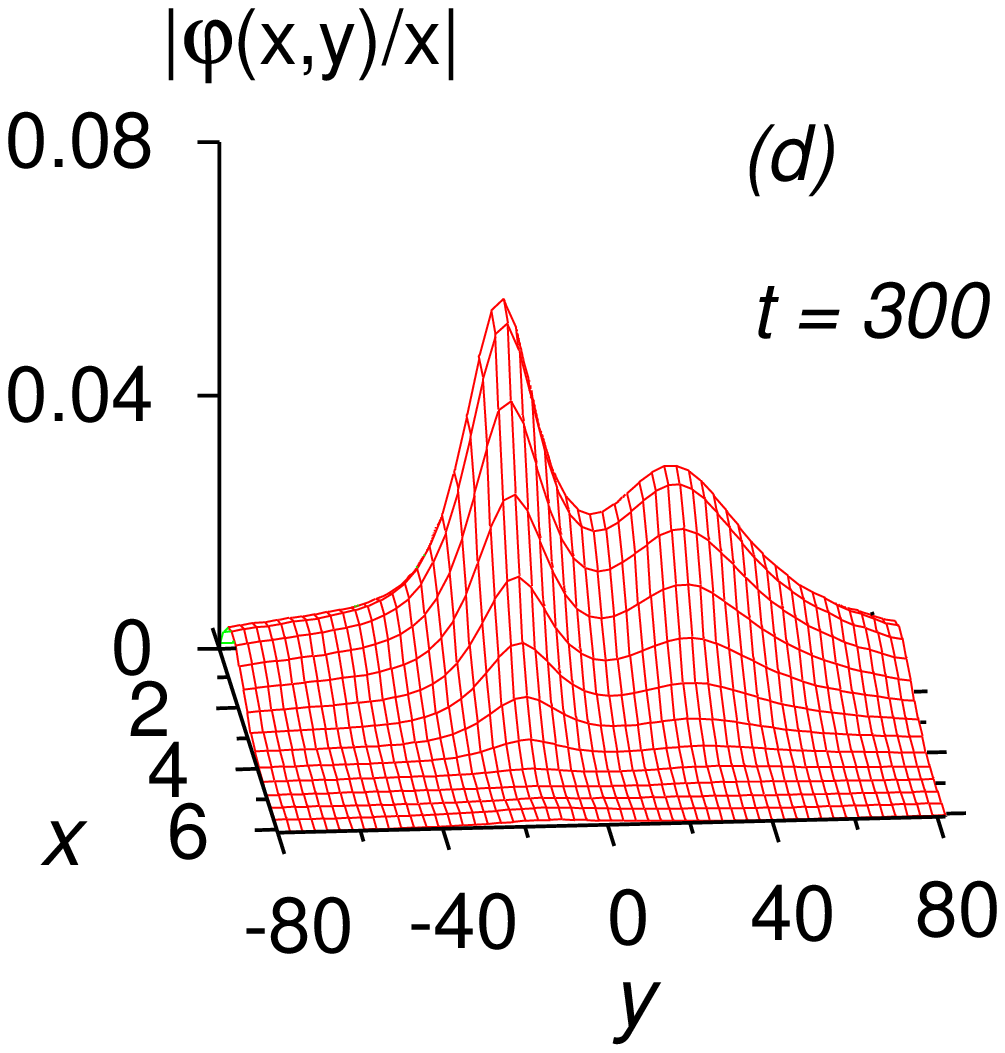}
\includegraphics[width=0.4\linewidth]{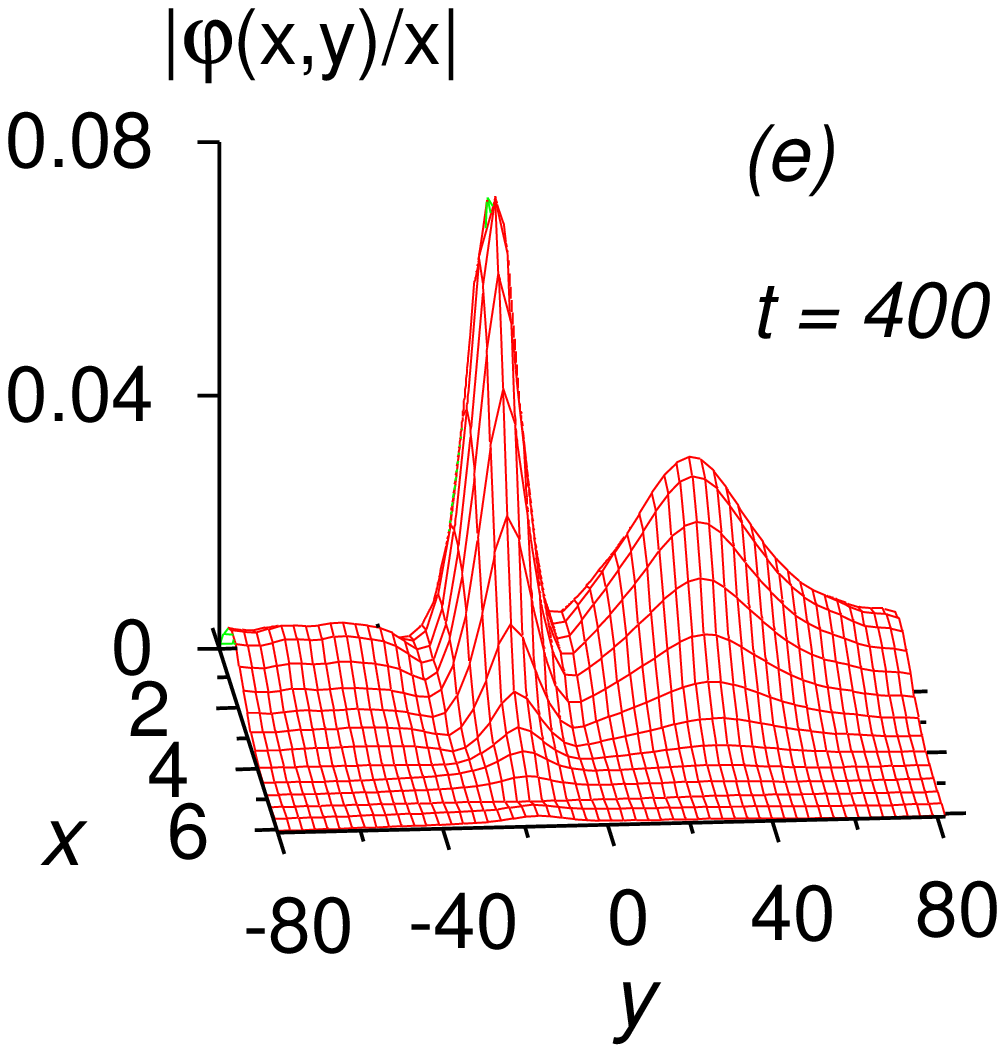}
\end{center}

\caption{The the-dimensional  wave function profile $|\varphi(x,y)/x|$
vs. $x$ and $y$
of two equal $L=1$ rotating bright solitons of figure 3 (b)  
with $n=-0.4$ each and $\delta
=\pi/4$ at times (a) $t=0$, (b) $t=100$, (c) $t =200$, (d) $t=300$,
and (e) $t=400$.
}
 
\end{figure}

\begin{figure}[!ht]
 
\begin{center}
\includegraphics[width=0.7\linewidth]{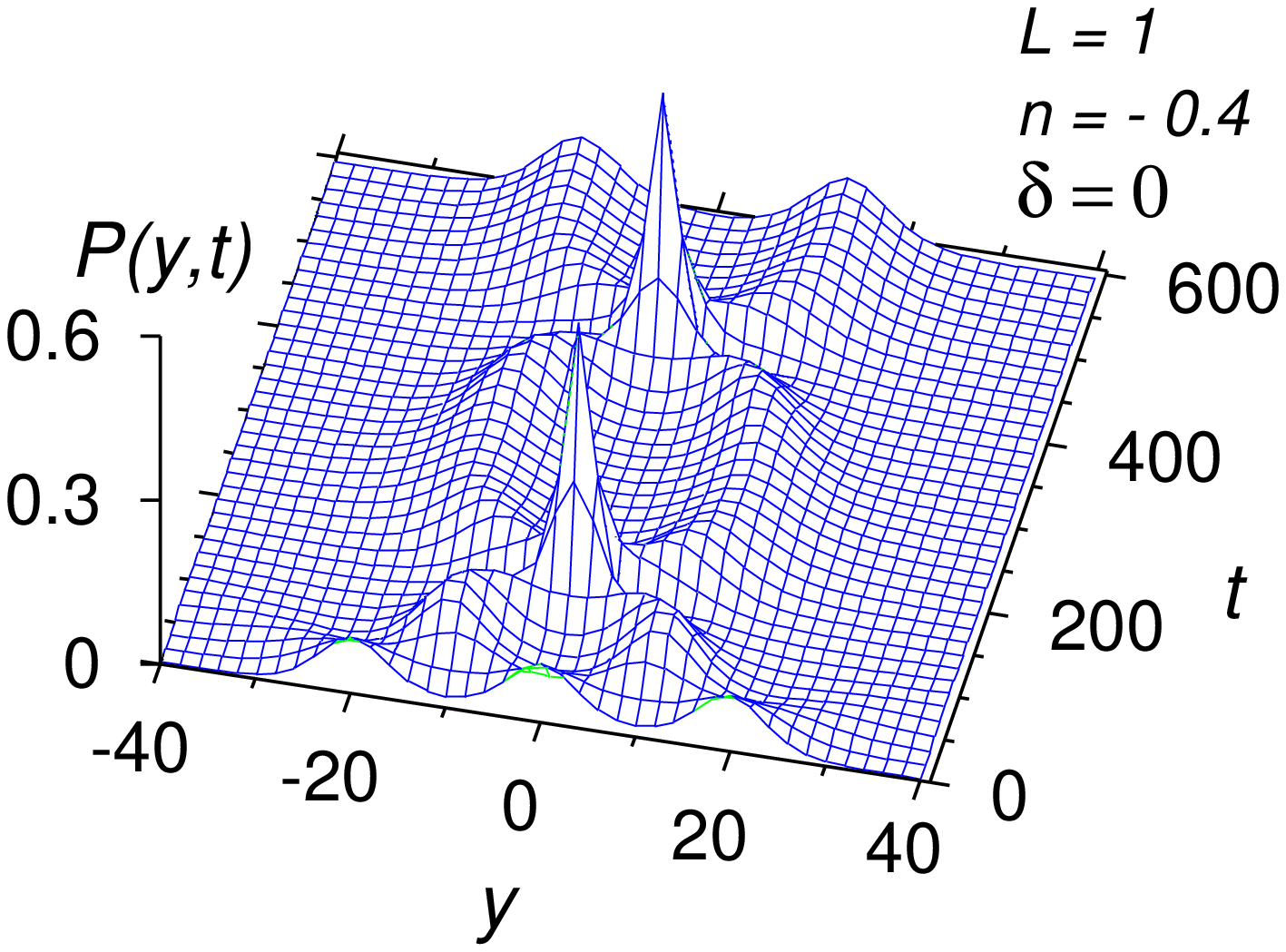}
\end{center}

\caption{One-dimensional probability  $P(y,t)$
vs. $y$ and $t$ for a train of three vortex solitons at 0 and $\pm 20$ 
each 
with   $n=-0.4,
L=1$ and   $\delta =0$.}

\end{figure}

Next we exhibit the profile of  two interacting solitons undergoing 
particle exchange. From figure 4 we find that $\delta = \pi/4$ leads to a
large exchange of particles. For this purpose we consider the situation
depicted in figure 3 (b) 
corresponding to two $L=1$ vortex solitons of $n=-0.4$
each which are set at $y= \pm 15$ at time $t=0$ with a phase difference of
$\delta = \pi/4$ according to (\ref{xxx}). The profiles of the system at
times $t=0, 100,200,$ 300 and 400  are shown in figures 5 (a), (b),
(c), (d) and (e), respectively. Due to exchange
of
particles the left soliton  gradually becomes larger and larger
consistent
with figure 4. At $t=100$ the solitons come closer,  interact 
strongly by exchanging atoms and become wider and deformed. 
Then they gradually move apart and
become narrower. Finally, at $t=400$ they re-acquire Gaussian shapes.
However, at $t=400$ the right soliton is wider and shorter and
accommodates a
smaller number 
of atoms consistent with figures 3 (b) and 4. At that stage the 
left soliton is narrower and taller and contains a larger fraction  of the
atoms.

The interaction between more than two solitons with a general phase
$\delta $ among different pairs is much too complicated to be
studied exhaustively here. If the phase difference $\delta$ between two
neighboring solitons is not close to zero, they experience overall
repulsion  and
stay apart. However, for $\delta$ close to zero they interact attractively
and often a soliton could be lost as observed in the experiment of 
Strecker {\it et al.} \cite{3}. This is illustrated in figure 6 where we
consider three solitons all with same
phase ($\delta = 0$). 
Due to attraction the three solitons come closer, interact
and form two solitons which after some time form a single soliton. 
This single soliton
next decays to two, recombines to one, and decays to two again  and never
three original solitons
are
recovered. It is also possible that for some values of nonlinearity $n$
and initial separation $2y_0$ two solitons with relative phase $\delta=0$ 
may   coalesce to form a single
soliton without ever decaying to two solitons again. These could explain a
missing soliton observed in the experiment by   Strecker {\it et al.}
\cite{3}. 

Throughout this investigation in the interaction of two equal solitons we
assumed that the nonlinearity $|n|$ for each is less than
$|n_{\mbox{cr}}|/2$, so that a stable solitonic condensate with total $|n|
<|n_{\mbox{cr}}|$ exists when the two  coalesce. However, if two solitons
each with  $|n| > |n_{\mbox{cr}}|/2$ encounter for $\delta =0$, the system
is expected to  coalesce, 
collapse and emit atoms via three-body recombination \cite{12}. It is
possible that in this case only a smaller single soliton survives.  This
might
also explain some  missing soliton(s) in experiment.

In the  investigation  by   Al Khawaja
{\it et
al.} \cite{6}, to justify the repulsion between solitons in a train
\cite{3} a
phase difference of $\delta = \pi$ was suggested between neighbors. 
They  made a
model for the evolution of  $\delta $ along the axial direction which they
used for  explaining the formation of a soliton train, with
neighboring solitons of phase difference $\pi$,
after changing
the scattering length from repulsive to attractive in a BEC as in the
experiment \cite{3}. Such an order in phase seems to be not  necessary for
an overall repulsion
between solitons. Almost any $\delta$, except $\delta$ near 0, is found to
lead to overall repulsion. In reference \cite{6a0}, using an approximate
analysis,
it has been found that  repulsive interactions require  $\cos \delta <0$.
In the present three-dimensional analysis, we find that attraction starts
at $\delta = \pi/2$ and increases as $\delta $ is reduced. However, the
overall interaction remains repulsive until a small value of $\delta$
close to zero is attained, when  the solitons attract, lose their separate
identity
and coalesce.

\section{Conclusion}

To conclude, employing numerical solution of the GP equation with axial
symmetry, we have performed a realistic mean-field study of interaction
among two vortex solitons in a train and find the overall interaction to
be repulsive except for phase $\delta $ between neighbors close to 0.  
For phase $\delta=\pi$ between neighboring solitons in a train, they are
found to repel and move away without exchanging atoms for both normal
($L=0$)  and vortex ($L=1$) solitons. For phase $\delta =0$ in a
two-soliton train, the solitons attract, collide, coalesce, and eventually
come out. For other $\delta$, overall repulsion prevails. However, there
is an inelastic exchange of atoms between two solitons resulting in a
change of size and shape. These unequal solitons travel in general with
different speeds: the smaller soliton travels faster and the larger
soliton travels slower. The unequal speed leads to an asymmetry in final
positions of the solitons. By changing the sign of the phase between the
two solitons the asymmetry in the final size and shape of the two
solitons can be reversed. Except in the $\delta \approx 0$ case, the
solitons in a train stay apart and never cross each other as observed in
the experiment by Strecker {\it et al.} \cite{3}.  For $\delta \approx 0$
a single soliton can often disappear as a result of the attractive
interaction among solitons, as observed experimentally by Strecker {\it et
al.} \cite{3}. These features of soliton interaction are present for both
normal ($L=0$) \cite{sala}  vortex ($L=1$) solitons.
 Although the present study is performed in the absence of
an axial trap as in one dimension \cite{1}, these general conclusions
should remain valid for a weak axial trap as in the experiments
\cite{3,4}.  The $L=1$ vortex solitons can accommodate larger number of
atoms and the present study may motivate future experiments with them.

\ack

I thank  Dr. F. Kh. Abdullaev,  Dr. A Gammal, Dr. Boris A. Malomed,
and Dr. Paulsamy
Muruganandam
for helpful discussions. 
The work was supported in part by the CNPq 
of Brazil.


\section*{References}
 
\begin{thebibliography}{99}

\bibitem{0}
 Hasegawa A and  Tappert F 1773 {\it Appl. Phys. Lett.} {\bf 23} 171

 Hasegawa A and  Tappert F 1773 {\it Appl. Phys. Lett.} {\bf 23} 142


Zakharov  V E and Shabat A B 1972 {\it Sov. Phys. JETP }  {\bf 34}  62

Zakharov  V E and Shabat A B 1973 {\it Sov. Phys. JETP }  {\bf 37}  823

 
\bibitem{1}Taylor J R 1992  Editor, {\it Optical Solitons - Theory and
Experiment} (Cambridge: Cambridge)



\bibitem{2} Denschlag J, Simsarian J E, Feder D L, Clark C W, Collins L A,
Cubizolles J, Deng L, Hagley E W, Helmerson K, Reinhardt W P, Rolston S L,
                              Schneider B I and  Phillips W D 2000
{\it  Science}  {\bf 287} 97


Anderson B P, Haljan P C, Regal C A, Feder D L, Collins L A, Clark C W and 
Cornell
E A 2001
  {\it Phys. Rev. Lett.} {\bf 86} 2926

Burger S, Bongs K, Dettmer S, Ertmer W, Sengstock K, Sanpera A,
Shlyapnikov G V and Lewenstein M
1999
   {\it Phys. Rev. Lett.}  {\bf 83} 5198

\bibitem{3}Strecker K E, Partridge G B, Truscott A G and Hulet R G 2002
{\it  Nature}  {\bf 417} 150

\bibitem{4} Khaykovich L, Schreck F, Ferrari G, Bourdel T, Cubizolles J,
Carr L D, Castin Y and Salomon C 2002 {\it  Science} {\bf 296} 1290


\bibitem{4a0} P\'erez-Garc\'ia V M,  Michinel H and  Herrero H
1998 {\it Phys. Rev. A} {\bf 57} 3837 


\bibitem{4a1}Muryshev A E,  van Linden van den Heuvell H B and
 Shlyapnikov G V 1999
 {\it   Phys. Rev. A} {\bf 60} R2665

\bibitem{12}Donley E A, Claussen N R, Cornish S L, Roberts J L, Cornell E
A and Wieman C E 2001  {\it Nature} {\bf 412} 295

Savage C M, Robins N P and  Hope J J 2003  {\it Phys. Rev. A} {\bf  67}
 014304 

Saito H and Ueda M  2002  {\it Phys. Rev. A} {\bf  65} 033624 

 Duine R A and  Stoof H T C 2003  {\it Phys. Rev. A} {\bf  68}  013602 



  Adhikari S K 2002 {\it Phys. Rev. A} {\bf  66} 043601

  Adhikari S K 2002 {\it Phys. Rev.  A} {\bf  66} 013611  


Fu H X, Li M Z, Gao B, Zhou S and  Wang Y Z 2002 \PL A {\bf 305}  204


\bibitem{4a}
 Inouye S, Andrews M R, Stenger J, Miesner H-J,  Stamper-Kurn D M  and
Ketterle W 1998 {\it Nature } {\bf 392} 151 

 Stenger J,  Inouye S,  Andrews M R,  Miesner H-J,
 Stamper-Kurn D M
and  Ketterle W 1999
 {\it  Phys. Rev. Lett. } {\bf 82} 2422

\bibitem{quote}   Hasegawa A 1984 {\it Opt. Lett. }  {\bf 9} 288

\bibitem{6}Al Khawaja U, Stoof H T C, Hulet R G, Strecker K E and
Partridge G B 2002 {\it Phys. Rev. Lett.} {\bf 89} 200404




\bibitem{6a0} Salasnich L, Parola A and Reatto L 2002 {\it Phys. Rev. A}
{\bf 66} 043603



\bibitem{leun}Leung V Y F, Truscott A G and Baldwin K G H 2002 {\it
Phys. Rev. A}
{\bf 66} 061602

\bibitem{sala}
 Salasnich L, Parola A and Reatto L  2003 \PRL {\bf 91} 080405


\bibitem{6a1}  Carr L D and  Castin Y 2003 {\it Phys. Rev. A}   {\bf 66}
063602



\bibitem{7x}Elyutin P V,  Buryak A V, 
 Gubernov V V,  Sammut R A and  Towers I N 
2001 {\it  Phys. Rev. E} {\bf 64} 016607

Shchesnovich   V S, Malomed B A and Kraenkel R A
2003 {\it Physica D} (in press) [{\it e-print nlin.PS/0210022}]


\bibitem{8}  Dalfovo F,  Giorgini S,  Pitaevskii L P and
 Stringari S 1999 {\it
Rev. Mod.  Phys.} {\bf 71} 463 


\bibitem{6a}                        Abo-Shaeer J R, Raman C, Vogels J M
and Ketterle W 2001
{\it  Science} {\bf 292} 476


Madison K W,  Chevy F,  Wohlleben W and  Dalibard J
2000   {\it   Phys. Rev. Lett.} {\bf 84} 806


Madison K W,  Chevy F, Bretin V  and  Dalibard J     2001   {\it
Phys. Rev. Lett.} 
 {\bf 86} 4443 


Matthews M R,  Anderson B P,  Haljan P C,  Hall D S,
 Wieman C E
and  Cornell E A
1999 {\it
Phys. Rev. Lett.}
{\bf
83}
2498 

\bibitem{vs} Dalfovo F and Stringari 1996 \PR A {\bf 53} 2477

 Dalfovo F  and Modugno M 2000 \PR A {\bf 61} 023605

Adhikari S K  2002 \PR A {\bf 65} 033616


\bibitem{9} Adhikari S K 2002 {\it Phys. Rev. E} {\bf 65}   016703


\bibitem{7}   Kivshar Y S, Alexander T J and  Turitsyn S K 2001  {\it 
 Phys. Lett. A}  {\bf 278} 225


\bibitem{5}  Desem C and Chu P L, Chapter 5 of  \cite{1}.










\bibitem{11} 
Adhikari S K and  Muruganandam P 2002 {\it
J. Phys. B: At. Mol. Opt. Phys.} {\bf 35} 2831

Adhikari S K and  Muruganandam P 2003 {\it
J. Phys. B: At. Mol. Opt. Phys.} {\bf 36} 409

  Muruganandam P  and  Adhikari S K  2003 \jpb
{\bf 36} 2501

\bibitem{10}Gerton J M, Strekalov D, Prodan I and Hulet R G 2001 {\it 
Nature } {\bf 408}   692 


\bibitem{crit}Gammal A, Tomio L and Frederico T 2002 {\it Phys. Rev.  A}
{\bf  66} 043619



\bibitem{gord}  Gordon J P 1983 {\it Opt. Lett.} {\bf 8} 596

 
 
 
 
 
 
 
 
 
\end{thebibliography}
\end{document}